\documentclass[reprint,
superscriptaddress,
amsmath,amssymb,amsart
aps,
floatfix,
longbibliography]{revtex4-1}

\usepackage{anyfontsize} %

\usepackage{colortbl}   
\usepackage{xcolor}     
\usepackage{amsmath,amssymb}
\usepackage{graphicx}
\usepackage{bm}

\usepackage{steinmetz}
\graphicspath{{./images/}}

\begin{document}
\title{Dynamics of amorphous membranes in the two-dimensional limit}

\author{Liga Jasulaneca}
\email{Correspondence to: Makars \v{S}i\v{s}kins (m.siskins@soton.ac.uk), Liga Jasulaneca (liga.jasulaneca@cfi.lu.lv)}
\affiliation{Institute of Solid State Physics, University of Latvia, Kengaraga Street 8, LV-1063, Latvia}
\author{Alberto Mart\'{\i}n-P\'erez}
\affiliation{Delft University of Technology, Mekelweg 2, Delft, 2628 CD, the Netherlands}
\author{Hongji Zhang}
\affiliation{Department of Materials Science and Engineering, National University of Singapore, Singapore}
\affiliation{Department of Physics, National University of Singapore, Singapore}
\author{Prertahn Munireternam}
\affiliation{Department of Materials Science and Engineering, National University of Singapore, Singapore}
\author{Natalia A. Mamchik}
\affiliation{Department of Materials Science and Engineering, National University of Singapore, Singapore}
\author{Chee-Tat Toh}
\affiliation{Department of Materials Science and Engineering, National University of Singapore, Singapore}
\affiliation{Department of Physics, National University of Singapore, Singapore}
\affiliation{Centre for Advanced 2D Materials Singapore, National University of Singapore, Singapore}
\author{Artem K. Grebenko}
\affiliation{Department of Materials Science and Engineering, National University of Singapore, Singapore}
\affiliation{Department of Physics, National University of Singapore, Singapore}
\author{Barbaros \"Ozyilmaz}
\affiliation{Department of Materials Science and Engineering, National University of Singapore, Singapore}
\affiliation{Department of Physics, National University of Singapore, Singapore}
\affiliation{Centre for Advanced 2D Materials Singapore, National University of Singapore, Singapore}
\affiliation{Institute for Functional Intelligent Materials, National University of Singapore, Singapore}
\author{Makars \v{S}i\v{s}kins}
\email{Correspondence to: Makars \v{S}i\v{s}kins (m.siskins@soton.ac.uk), Liga Jasulaneca (liga.jasulaneca@cfi.lu.lv)}
\affiliation{School of Physics and Astronomy, University of Southampton, Southampton, SO17 1BJ, United Kingdom}
\author{Farbod Alijani}
\affiliation{Delft University of Technology, Mekelweg 2, Delft, 2628 CD, the Netherlands}


\begin{abstract}Atomically thin mechanical resonators have been realized predominantly in crystalline two-dimensional (2D) materials, such as graphene, where long-range crystalline order sets their elastic properties and defines their nonlinear resonant behavior. Extending these concepts to the amorphous 2D limit has remained largely unexplored. Here, we demonstrate that monolayer amorphous carbon (MAC) forms suspended membranes that support optothermal actuation and sensitive interferometric readout across both linear and nonlinear regimes of its resonant motion. We resolve thermomechanical motion, driven resonances, and multimode spectra in MAC nanodrums. The frequencies of fundamental vibration modes correspond to unusually low pretensions, placing monolayer MAC nanodrums in a regime where geometric nonlinearities, stress heterogeneity, and mode coupling emerge at comparatively low drive powers. Consistently, we observe pronounced nonlinear dynamics, including hardening, softening, and mixed Duffing responses, nonlinear damping, parametrically excited modes, and signatures of intermodal coupling. These results establish MAC as a robust nanoelectromechanical platform and open an experimental route to disorder-governed nanomechanics in the 2D amorphous limit.
\end{abstract}

\maketitle

\section*{Introduction}\label{sec1}

Atomically thin suspended membranes of two-dimensional (2D) materials combine low mass, strong in-plane covalent bonding, and extreme out-of-plane flexibility, providing access to resonant mechanical motion across a wide dynamic range \cite{lemme_nanoelectromechanical_2020_fix, steeneken_dynamics_2021}. In these materials, thermally driven Brownian motion gives access to intrinsic properties such as effective mass, tension and dissipation, while externally driven motion reveals nonlinear phenomena including amplitude-dependent frequency shifts, nonlinear damping and mode interactions \cite{davidovikj_nonlinear_2017,keskekler_tuning_2021,kaisar_nonlinear_2022}. These capabilities have established 2D resonators as versatile systems for nanoscale materials characterization \cite{davidovikj_nonlinear_2017, Wrinkles_ArjmandiTash2026}, tunable sensing \cite{liu_enhanced_2024, kartal_frequency_2026_fix} and studies of coupling between mechanical motion and electronic, optical, magnetic and thermal degrees of freedom \cite{keskekler_tuning_2021,kaisar_nonlinear_2022, zheng_hexagonal_2017, siskins_magnetic_2020,liu_tuning_2023_fix}. A range of crystalline 2D materials are utilized as suspended resonators \cite{keskekler_tuning_2021,kaisar_nonlinear_2022, zheng_hexagonal_2017, siskins_magnetic_2020,liu_tuning_2023_fix}, including graphene \cite{keskekler_tuning_2021}, MoS$_2$ \cite{kaisar_nonlinear_2022}, hBN \cite{zheng_hexagonal_2017}, and FePS$_3$ \cite{siskins_magnetic_2020}. So far, such studies have largely focused on crystalline membranes, whose elastic behavior is closely linked to lattice symmetry and long-range order. 

\begin{figure*}
\centering
\includegraphics[width=\textwidth]{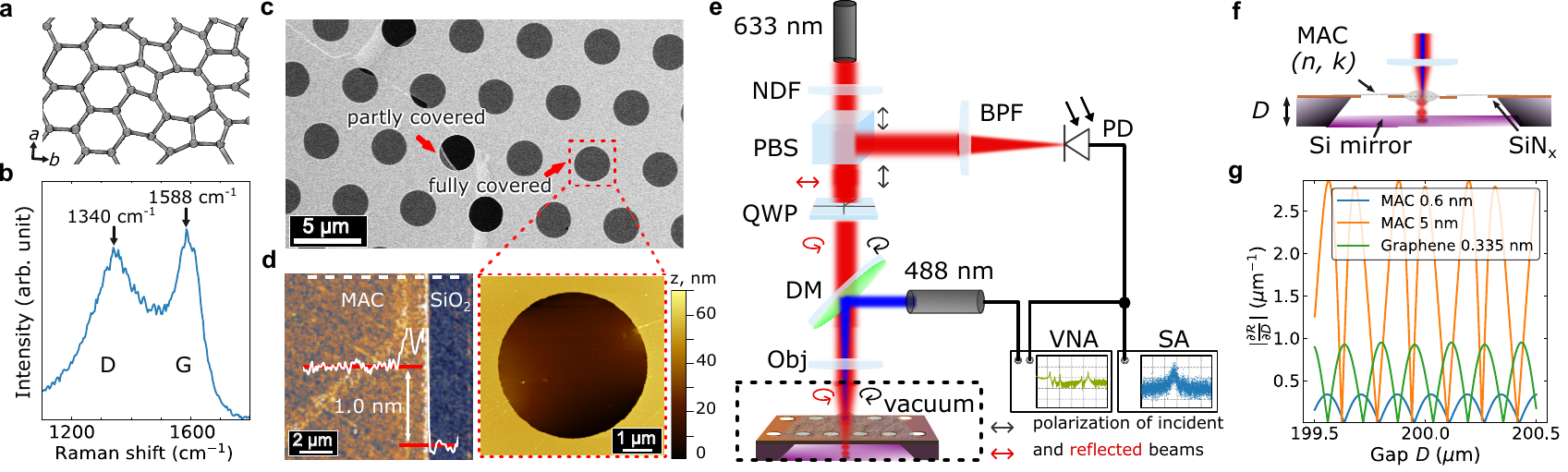}
\caption{Optothermal actuation and interferometric readout of vibrating monolayer amorphous carbon (MAC) membranes.
(a) Schematic illustration of MAC, highlighting its structural characteristics: five-, six-, seven-, and eight-member rings, varying bond angles and lengths, and locally hexagonal patches.
(b) Raman spectrum acquired on suspended monolayer MAC, showing the disorder-activated D band near 1340~cm$^{-1}$ and the G band near 1588~cm$^{-1}$, consistent with disordered $\mathrm{sp}^2$ carbon.
(c) SEM image of MAC suspended over a Si-supported holey SiN$_x$ array, showing continuous suspended regions across the circular openings.
(d) AFM characterization of transferred MAC. Left: AFM topography and height profile of MAC on SiO$_2$, showing an approximately 1~nm step height consistent with an atomically thin transferred film. Right: AFM image of an intact suspended MAC nanodrum.
(e) Schematic of the actuation and readout setup. Blue and red beams were focused onto the membrane through a $50\times$ objective (NA~=~0.42). NDF, neutral density filter; PBS, polarizing beam splitter; QWP, quarter-wave plate; DM, dichroic mirror; BPF, bandpass filter; PD, photodetector; VNA, vector network analyzer; SA, spectrum analyzer.
(f) Schematic of the Fabry--P\'{e}rot-type configuration formed by the suspended MAC membrane and the underlying Si substrate.
(g) Calculated power-normalized interferometric responsivity $\left|\partial\mathcal{R}/\partial D\right|$ as a function of membrane--substrate gap for monolayer MAC, 5~nm MAC, and monolayer graphene.}
\label{fig1}
\end{figure*}

Amorphous 2D resonators, in contrast, remain much less explored. Thicker amorphous nanomechanical resonators such as silicon nitride \cite{tsaturyan_ultracoherent_2017} and silicon carbide \cite{xu_high-strength_2024} exhibit high mechanical stability and quality factors despite the absence of crystalline order, demonstrating that long-range periodicity is not required for high-performance nanoelectromechanical systems. However, realizing freestanding membrane resonators made of amorphous materials in the 2D limit remains challenging, because the membrane must remain continuous, survive transfer, and form stable suspended structures over micron-scale cavities. 
Approaches for creating 2D amorphous resonators include laser oxidation of crystalline 2D materials~\cite{cartamil-bueno_high-quality-factor_2015} and electron-induced cross-linking of aromatic self-assembled monolayers to form carbon nanomembranes (CNMs)~\cite{zhang_vibrational_2015}. CNMs represent an important example of ultrathin amorphous resonators, with vibrational modes described within a linear membrane model~\cite{zhang_vibrational_2015}. However, their formation through cross-linking of molecular precursors imposes limitations on synthesis and subsequent single-layer manipulation and transfer onto substrates. Monolayer amorphous carbon (MAC) provides a distinct route to the atomically thin amorphous limit, combining large-area continuity and mechanical robustness in a single freestanding membrane. It is synthesized as a free-standing monolayer comprising an $\mathrm{sp}^2$-bonded carbon network that is amorphous at the atomic scale~\cite{toh_synthesis_2020}. Its lack of in-plane long-range order makes it a model system for studying how structural heterogeneity shapes electrical and mechanical properties in a 2D membrane~\cite{tian_disorder-tuned_2023}. Additionally, in contrast to graphene, MAC can be reliably grown wide-bandgap and electrically insulating~\cite{toh_synthesis_2020}, suppressing charge transport and enabling mechanical dynamics to be probed in relative isolation from damping effects of electronic origin~\cite{will_high_2017}. These properties make MAC a particularly attractive candidate for studying nanomechanics in an amorphous monolayer. 

Here, we realize above-bandgap optothermal actuation and below-bandgap interferometric readout to characterize the electrically insulating membranes of atomically thin layers of MAC. We combine micron-scale mechanically intact suspended MAC nanodrums with the optical actuation and readout scheme to characterize both its linear and amplitude-dependent vibrational response. The measured dynamics is interpreted using the tensioned membrane description consistent with that applied to 2D nanodrums. Within this framework, monolayer MAC exhibits low effective pretension, suggesting a regime where the effects of membrane heterogeneity are expected to be particularly pronounced in multimode and nonlinear dynamics.

\section*{Results and Discussion}\label{sec2}

We synthesized atomically thin MAC by a laser-assisted chemical vapor deposition method~\cite{toh_synthesis_2020}. Fig.~\ref{fig1}(a) schematically illustrates its disordered $\mathrm{sp}^2$-bonded network, including non-hexagonal rings, variations in bond lengths and angles, and locally hexagonal regions. Raman spectra acquired on suspended membranes show the characteristic disorder-activated D and G bands of $\mathrm{sp}^2$ carbon (Fig.~\ref{fig1}(b)), consistent with the Raman fingerprint of MAC \cite{toh_synthesis_2020}. Tapping-mode atomic force microscopy (AFM) of MAC transferred onto SiO$_2$ yielded an step height of approximately 1~nm, consistent with an atomically thin transferred film \cite{MonolayerMAC_Zhang2026}, taking into account scanning probe and interface-related offsets \cite{Thickness_Shearer2016} (Fig.~\ref{fig1}(d)). Resonators were fabricated by transferring monolayer and few-layer ($\sim$5~nm) MAC onto SiN$_x$/Si (0.1/200~$\mu$m) transmission electron microscopy grids containing circular through-holes with radii 1.25~$\mu$m and 2.5~$\mu$m using PMMA-assisted wet transfer~\cite{toh_synthesis_2020} (Fig.~\ref{fig1}(c)). The through-hole geometry provided a reliable suspension yield, and, together with the underlying silicon surface, defined a Fabry--P\'{e}rot (FP)-type optical cavity. AFM measurements showed relatively flat suspended membranes recessed approximately 40--100~nm below the top SiN$_x$ surface, likely because of adhesion to the hole sidewalls (Fig.~\ref{fig1}(d)). 

\begin{figure} [t]
\centering
\includegraphics[width=\linewidth]{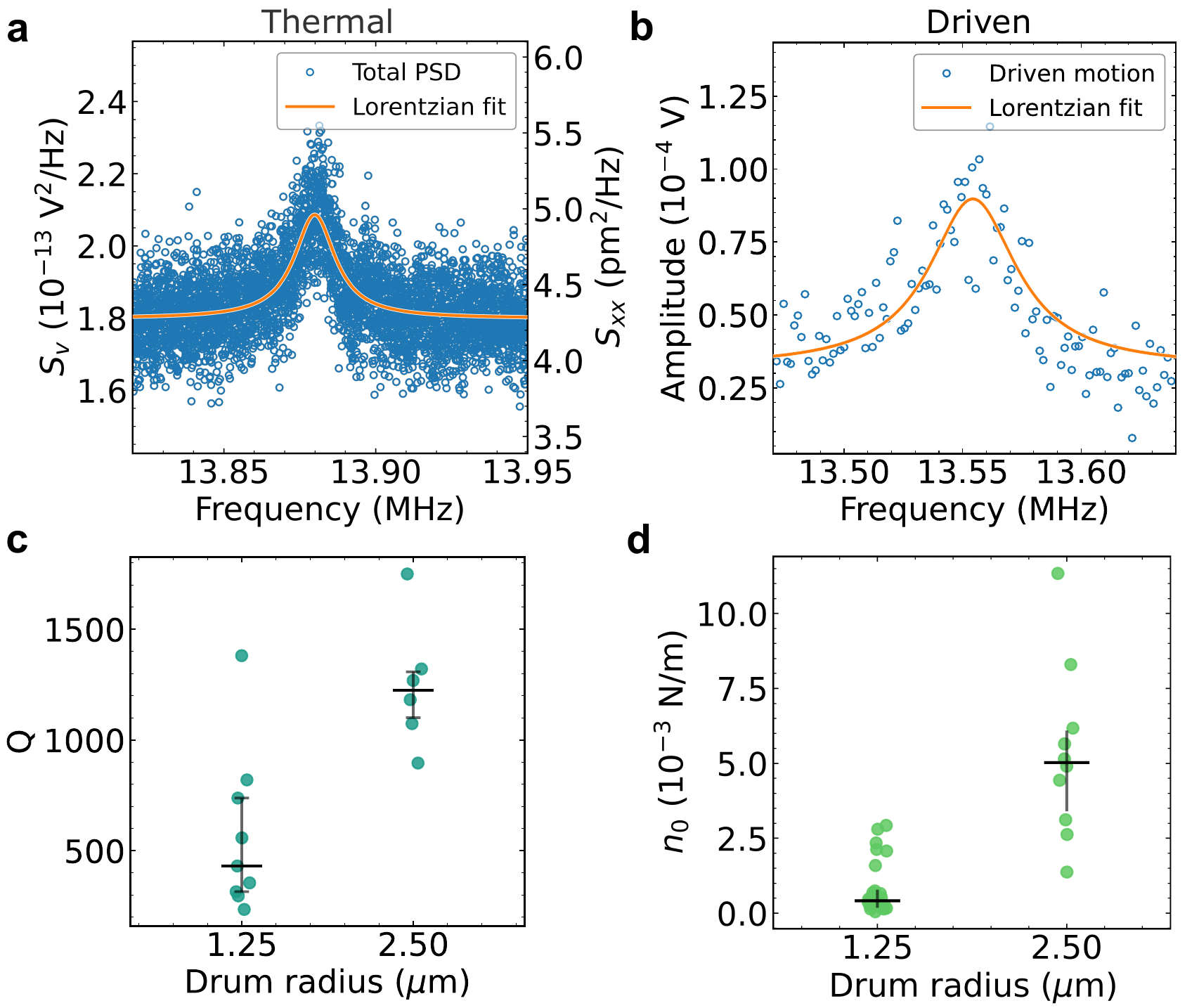}
\caption{Linear dynamic response of monolayer MAC nanodrums.
(a) Power spectrum showing a Lorentzian-shaped resonance from which a fundamental frequency of 13.9~MHz and $Q_{\mathrm{th}} = 843 \pm 24$ were obtained.
(b) Driven fundamental mode of the same nanodrum, with $f_{0,\mathrm{driven}}\approx13.6$~MHz and $Q_{\mathrm{driven}}=302\pm 27$.
(c) $Q$ distribution for MAC nanodrums with radius $R = 1.25~\mu$m and $2.50~\mu$m. 
(d) Pretension distributions for the same radius values. 
Horizontal and vertical black bars indicate the median values and interquartile ranges of $n_0$ and $Q$. Data points are slightly scattered in radius value for clarity.}
\label{fig:lin}
\end{figure}

The dynamical characterization of these membranes was carried out in a vacuum chamber at a base pressure of $10^{-6}$~mbar. Membrane motion was actuated optothermally using a power-modulated 488~nm diode laser (Fig.~\ref{fig1}(e)). The motion was read out interferometrically with a 633~nm He--Ne laser, following established approaches for 2D resonators \cite{davidovikj_visualizing_2016}. The time-averaged optical power at the membrane was $0.2$--$2.0$~mW for the modulated 488~nm excitation beam, while the 633~nm readout power was $400$~$\mu$W.

The displacement-to-optical-signal conversion was described using an FP-type interference model. In this geometry, the semi-transparent membrane, of thickness $h$, and the reflective silicon surface at the bottom of the through-hole are separated by a cavity depth $D$ of approximately 200~$\mu$m (Fig.~\ref{fig1}(f)). The reflected intensity is modulated by interference between light reflected from the membrane and the silicon substrate. At normal incidence, the Fresnel reflection coefficient of a single air--membrane interface is
\begin{equation}
r_{\mathrm{m}} = \frac{1-\tilde{n}}{1+\tilde{n}},
\end{equation}
where $\tilde{n}=n-ik$ is the complex refractive index, with real part $n$ and extinction coefficient $k$. For the device geometry, reflections at all interfaces and phase accumulation within the membrane and vacuum gap were included using a transfer-matrix model \cite{blake_making_2007}. The resulting complex reflection amplitude of the complete air--membrane--vacuum-gap--silicon stack is denoted by
$r_{\mathrm{tot}}(D)$, where $D$ is the membrane--substrate separation. The corresponding reflectance is 
\begin{equation}
\mathcal{R}(D)=\left|r_{\mathrm{tot}}(D)\right|^2.
\end{equation}
The reflectance varies approximately sinusoidally with $D$~\cite{dolleman_amplitude_2017}. The power-normalized displacement responsivity is $\left|\partial\mathcal{R}/\partial D\right|$ and quantifies the change in reflected intensity per unit membrane displacement and per unit incident intensity.

Monolayer MAC exhibits an optical bandgap around 2.1~eV~($\sim$590~nm), with an enhanced real part $\varepsilon_1$ and a reduced imaginary part $\varepsilon_2$ of the dielectric function near the absorption onset~\cite{toh_synthesis_2020}. At the 633~nm transduction wavelength, which lies below the bandgap, $\varepsilon_1 \approx 8.4$ and $\varepsilon_2 \approx 0.15$, estimated from ellipsometric data of as-synthesized MAC monolayer~\cite{toh_synthesis_2020}. Since the complex dielectric function and refractive index are related by $\varepsilon=\tilde{n}^2$, these values correspond to a high-$n$, low-$k$ regime. This provides predominantly dispersive transduction, while the much smaller extinction coefficient compared with monolayer graphene~\cite{wang_strong_2008} implies reduced optical absorption. In contrast, the actuation wavelength of 488~nm lies above the bandgap of MAC, where increased absorption enables efficient optothermal driving. The calculated power-normalized displacement responsivities $\left|\partial\mathcal{R}/\partial D\right|$ for monolayer and 5-nm-thick MAC in vacuum, together with monolayer graphene as a reference (Fig.~\ref{fig1}(g)), show that monolayer MAC provides a responsivity of the same order of magnitude as graphene in this transduction geometry. For 5-nm-thick MAC, the calculated responsivity exceeds that of monolayer graphene, indicating that modestly thicker MAC membranes can provide enhanced interferometric transduction while retaining low optical absorption at the readout wavelength.

To estimate the membrane displacement, we measured the power spectral density of a fundamental mode of a representative 2.5~$\mu$m-radius monolayer nanodrum using a spectrum analyzer, as shown in Fig.~\ref{fig:lin}(a). The resonance exhibits a Lorentzian lineshape characteristic of a damped harmonic oscillator. A Lorentzian fit with a constant background yielded a resonance frequency $f_{0,\mathrm{th}}\approx 13.9$~MHz and a quality factor $Q_{\mathrm{th}} = 843 \pm 24$.

By subtracting a frequency-independent background $S_{v,\mathrm{bg}}(f)$ from the measured voltage noise spectral density $S_v(f)$, and integrating the mechanical contribution over a frequency window around the resonance, the mean-square voltage fluctuations associated with Brownian motion were obtained as
\begin{equation}
\langle v^2 \rangle =
\int \left[S_v(f)-S_{v,\mathrm{bg}}(f)\right]\,df .
\end{equation}

The resonance frequency $f_0$ and linewidth $\gamma$ were extracted from Lorentzian fits with a fixed background level. The mean-square displacement was obtained from the equipartition theorem as
\begin{equation}
\langle x^2 \rangle =
\frac{k_{\mathrm{B}}T}{m_{\mathrm{eff}} (2\pi f_0)^2},
\end{equation}
where $T$ is the temperature and $m_{\mathrm{eff}}$ is the effective modal mass. $T$ was assumed to be equal to the room temperature of 293~K and the modal mass was estimated from the membrane geometry assuming a circular nanodrum with uniform areal mass density $\rho_{\mathrm{2D}}$,
\begin{equation}
m_{\mathrm{eff}} = \alpha \rho_{\mathrm{2D}} \pi R^2 ,
\end{equation}
where $\alpha = 0.269$ corresponds to the fundamental mode of a tension-dominated circular membrane. Since the areal mass density of suspended monolayer MAC has not been directly measured, we approximate it as $\rho_{\mathrm{2D}}=\rho h$, using an effective density $\rho = 1300~\mathrm{kg\,m^{-3}}$ and nominal monolayer thickness $h = 0.6$~nm, appropriate for a predominantly threefold-coordinated carbon network~\cite{toh_synthesis_2020}.

\begin{figure*}
\centering
\includegraphics[width=0.8\textwidth]{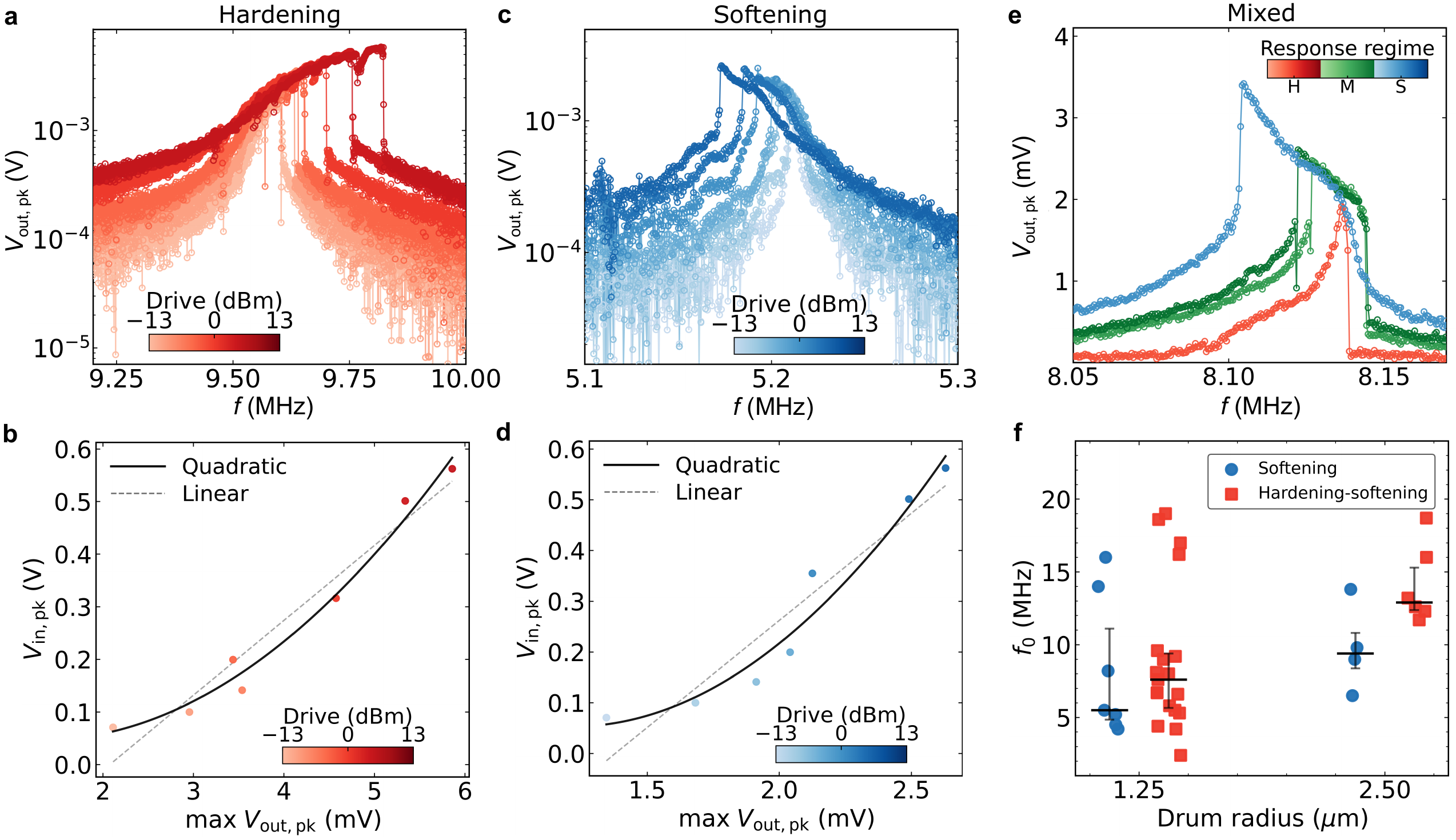}
\caption{Amplitude-dependent nonlinear dynamics of MAC resonators. Near-resonant driven responses showing hardening (a) and softening (c) Duffing-type nonlinearities with increasing blue laser drive power. (b, d) Nonlinear damping, indicated by the quadratic relationship between the drive and maximum response amplitudes. Dashed and solid lines show linear and quadratic fits, respectively. (e) Transition from hardening (H) to mixed (M) to softening (S) in a representative MAC nanodrum and (f) statistics of nanodrums showing different nonlinear response types.}
\label{fig:8panel2}
\end{figure*}

Comparing the measured voltage variance $\langle v^2\rangle$ with the thermal displacement variance $\langle x^2\rangle$ yields an estimate of the optical transduction factor in units of nm/V. The thermomechanical spectrum of the same 2.5~$\mu$m-radius monolayer MAC nanodrum was calibrated using the equipartition theorem, giving $x_{\mathrm{rms}}\approx0.4$~nm and an optical transduction factor of $(4.88\pm0.05)\times10^{3}$~nm/V. The calibrated displacement noise floor corresponds to a displacement sensitivity of $\sim2.1~\mathrm{pm}/\sqrt{\mathrm{Hz}}$, or equivalently a displacement power spectral density of $\sim4.4~\mathrm{pm^2/Hz}$. The Brownian resonance peak rises by approximately $1~\mathrm{pm^2/Hz}$ above this background. This sensitivity is comparable to the $0.6~\mathrm{pm}/\sqrt{\mathrm{Hz}}$ reported for graphene optomechanical resonators~\cite{barton_photothermal_2012}, despite the non-optimized interferometric geometry used here, with a mirror separation of approximately 200~$\mu$m. The calibration is limited mainly by uncertainty in modal mass, possible laser-induced temperature offsets, and deviations from an ideal tension-dominated mode shape. Across three devices with sufficient Brownian signal, the extracted transduction factors varied within a factor of 3. The resonance frequency showed only weak dependence on the readout laser power, in contrast to the linewidth.

The same device was subsequently actuated using amplitude-modulated 488~nm optothermal excitation. The driven response (Fig.~\ref{fig:lin}(b)) yielded $f_{0,\mathrm{driven}}\approx13.6$~MHz and $Q_{\mathrm{driven}} = 302\pm27$. Consequently, the difference in resonance frequency and quality factor between thermomechanical and driven measurements is attributed to optothermal heating introduced by the blue laser and to operation at higher amplitudes in the driven regime.

Across the measured MAC monolayers, the experimentally determined fundamental frequencies range from 2.4 to 19~MHz, with quality factors from 235 to 1750 (Fig.~\ref{fig:lin}(c)). Nanodrums with $R=2.5~\mu\mathrm{m}$ showed an approximately twofold higher median $Q$ than those with $R=1.25~\mu\mathrm{m}$. A similar diameter dependence has been reported for graphene membranes and attributed to geometry-dependent dissipation~\cite{barton_high_2011}.
We use the obtained resonance frequencies to estimate an equivalent two-dimensional pretension $n_0$ within the standard tension-dominated circular membrane model commonly used for suspended 2D material nanodrums~\cite{steeneken_dynamics_2021,davidovikj_nonlinear_2017}. In this model, the axisymmetric fundamental frequency is
\begin{equation}
f_0 = \frac{\alpha_{01}}{2\pi R}
\sqrt{\frac{n_0}{\rho_{\mathrm{2D}}}} ,
\end{equation}
where $\alpha_{01}=2.4048$ is the first zero of the zeroth-order Bessel function of the first kind, $J_0$, $R$ is the membrane radius, $n_0$ is the two-dimensional pretension, and $\rho_{\mathrm{2D}}$ is the areal mass density defined above~\cite{leissa_vibrations_2011}. Using this expression, the measured fundamental frequencies yield $n_0$ in the range $10^{-4}-10^{-3}$~N~m$^{-1}$ (Fig.~\ref{fig:lin}(d)). These values should be regarded as order-of-magnitude estimates of the effective pretension. Nevertheless, they are one to two orders of magnitude lower than the pretensions commonly reported for suspended crystalline graphene membranes \cite{bunch_impermeable_2008}, placing monolayer MAC in a low-effective-pretension regime. 
\begin{figure} [t]
\centering
\includegraphics[width=\linewidth]{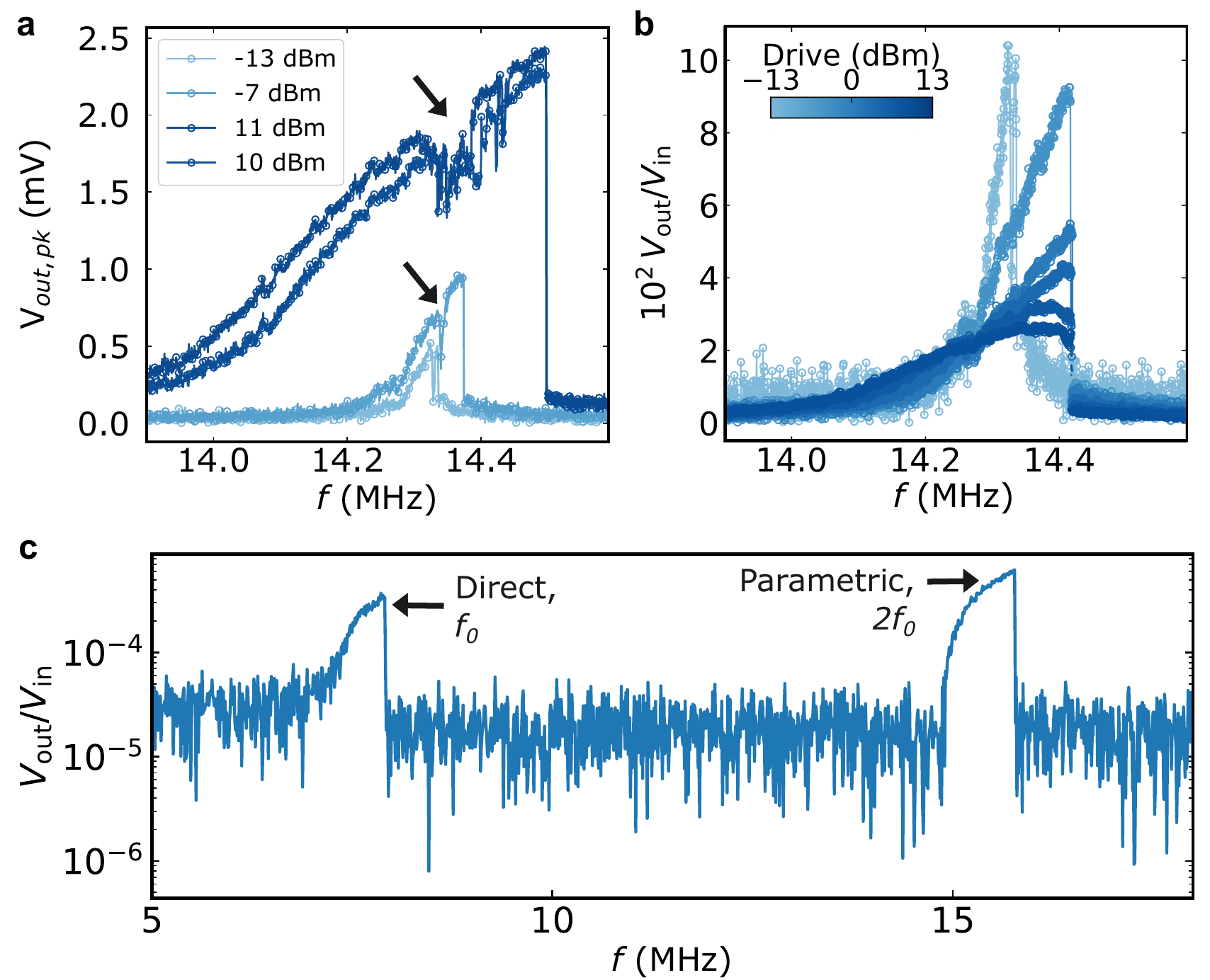}
\caption{Signatures of modal interactions in a MAC nanodrum. (a) Frequency response at increasing drive, showing the development of a dip in the resonance peak amplitude, consistent with the onset of internal resonance. (b) The same response plotted in normalized units, \(V_{\mathrm{out}}/V_{\mathrm{in}}\), highlights the strong reduction in the apparent quality factor \(Q\). (c) Mode spectrum of a similar MAC nanodrum, showing secondary mode near \(2f_0\), suggesting a \(2{:}1\) internal resonance.}
\label{fig:internal}
\end{figure}
This low-pretension regime is relevant not only for the linear resonance frequencies, but also for the expected onset of amplitude-dependent dynamics. 

Following the linear dynamics characterization, we further consider the large-amplitude response of MAC resonators at higher optothermal drive power. To describe this high-amplitude behavior, we model the fundamental flexural mode as a driven Duffing resonator with amplitude-dependent damping,
\begin{equation}
m_{\mathrm{eff}}\ddot{x}
+m_{\mathrm{eff}}(\gamma_1+\gamma_3 x^2)\dot{x}
+(k_1+k_3x^2)x
=F\cos(\omega t),
\end{equation}
where $m_{\mathrm{eff}}$ is the effective modal mass, $F$ and $\omega$ are the drive amplitude and angular frequency, and $k_1=m_{\mathrm{eff}}\omega_0^2$. The coefficient $k_3$ describes the cubic Duffing stiffness, while $\gamma_1=\omega_0/Q$ and $\gamma_3$ describe the linear and nonlinear damping, respectively. The onset of non-negligible nonlinear response may be estimated by taking the cubic stretching force to reach 10\% of the linear restoring force, giving an onset displacement that scales as~\cite{steeneken_dynamics_2021}
\begin{equation}
x_{\mathrm{nl}} \propto R\sqrt{\frac{n_0}{Eh}},
\end{equation}
where \(R\) is the drum radius, \(n_0\) is the two-dimensional pretension, \(E\) is Young's modulus, and \(h\) is the membrane thickness. This scaling indicates that, for devices of comparable radius and comparable two-dimensional stiffness \(Eh\), the one- to two-order-of-magnitude reduction in pretension inferred above would lower the expected nonlinear-onset displacement by approximately a factor of 3--10. 

Fig.~\ref{fig:8panel2} shows representative examples of amplitude-dependent responses observed in different MAC nanodrums. As the optothermal drive power was increased, a characteristic bending of the resonance curve toward higher frequencies was observed near resonance (Fig.~\ref{fig:8panel2}(a)). This behavior is the signature of Duffing hardening nonlinearity, corresponding to a positive cubic stiffness coefficient $k_3>0$. A number of other resonators exhibited a decrease of the resonance frequency with increasing drive power (Fig.~\ref{fig:8panel2}(c)), attributed to a softening nonlinearity corresponding to $k_3<0$. In atomically thin nanodrum resonators, such softening can arise from geometric asymmetries, uneven tension, membrane bulging, or motion-dependent optothermal effects \cite{kaisar_nonlinear_2022}. Additionally, nonlinear damping becomes apparent when the required drive amplitude is compared with the peak response amplitude. For both hardening and softening cases, $V_{\mathrm{in,pk}}$ exhibits an approximately quadratic dependence on $\max\left[V_{\mathrm{out,pk}}(f)\right]$ (Fig.~\ref{fig:8panel2}(b,d)). This behavior is consistent with amplitude-dependent nonlinear damping described by the $m_{\mathrm{eff}}\gamma_3 x^2 \dot{x}$ term in the equation of motion, as reported previously for 2D nanomechanical resonators at large vibration amplitudes~\cite{zhang_probing_2023,kartal_frequency_2026_fix}.

Notably, the nonlinear stiffness is not fixed in sign for all devices. In some nanodrums, an initial hardening response transitions into a mixed regime and then into softening behavior (Fig.~\ref{fig:8panel2}(e)). Such coexistence of hardening and softening suggests competing nonlinear contributions. Similar sign changes of the effective nonlinear stiffness have been reported when multiple nonlinear mechanisms coexist, including static curvature, symmetry breaking, higher-order nonlinearities, strain inhomogeneity, and motion-dependent laser-heating effects~\cite{samanta_tuning_2018, kaisar_nonlinear_2022,arora_mixed_2024,li_strain_2024}. Across the full monolayer dataset, these responses fall mainly into two categories: purely softening behavior and mixed hardening--softening behavior (Fig.~\ref{fig:8panel2}(f)). The occurrence of pure softening correlates with lower fundamental resonance frequencies, consistent with a stronger influence of low pretension, static deformation, or spatially nonuniform stress in the lowest-frequency devices.

\begin{figure*}
\centering
\includegraphics[width=\textwidth]{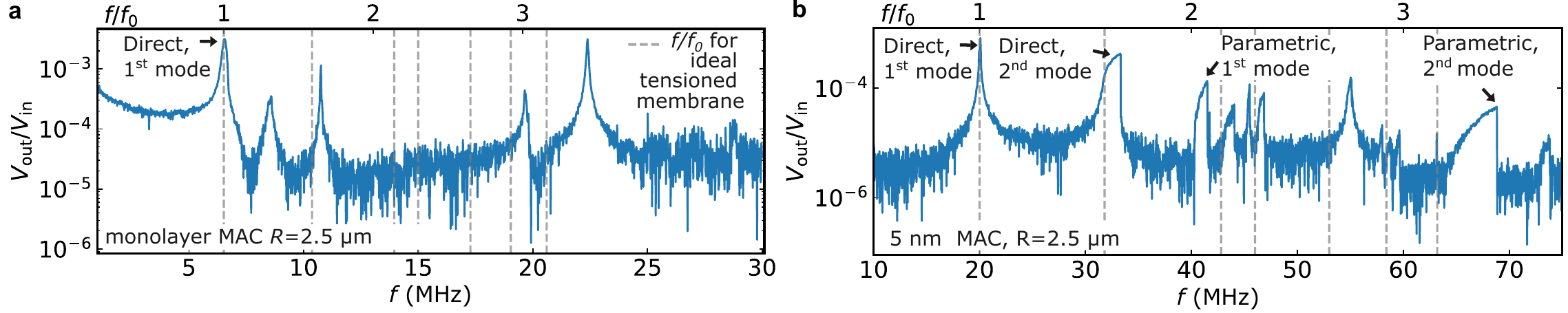}
\caption{Mode spectra of monolayer and multilayer MAC nanodrums. Representative spectra of (a) monolayer and (b) multilayer MAC nanodrums show multiple vibrational modes and nonlinear response features. Thin vertical lines are guides to the eye marking the mode-frequency ratios expected for an ideal tension-dominated circular membrane.}
\label{fig:mode}
\end{figure*}

Furthermore, a set of membrane devices shows signatures of dynamics beyond a single-mode Duffing response (Fig.~\ref{fig:internal}). In the frequency-response curves, the resonance peak develops a pronounced dip that remains visible over several drive levels~(Fig.~\ref{fig:internal}(a)). When plotted in drive-normalized units, $V_{\mathrm{out}}/V_{\mathrm{in}}$~(Fig.~\ref{fig:internal}(b)), the response shows a threshold-like reduction in the apparent quality factor, with an abrupt drop occurring at nearly the same frequency for successive drive amplitudes~\cite{keskekler_tuning_2021,houri_demonstration_2020} (Fig.~\ref{fig:internal}(b)). This behavior points to dynamics beyond a single-mode hardening Duffing response, for which the resonance maximum would generally be expected to continue shifting upward with increasing drive.  Instead, the dip and associated reduction in apparent $Q$ suggest an additional amplitude-dependent dissipation channel. A plausible mechanism is intermodal energy transfer mediated by internal resonance, where nonlinear frequency tuning brings two modes into a near-commensurate relation and enables efficient energy exchange. Similar behavior has been reported in graphene nanodrums, where parametric--direct $2{:}1$ internal resonance enhances effective nonlinear damping, and in MoS$_2$ nanodrum resonators, where enhanced nonlinear damping appears in the parameter space favoring internal resonance \cite{keskekler_tuning_2021,prasad_tunable_2023}. Consistent with this interpretation, spectra acquired at the same drive level show a secondary mode near $2f_0$ (Fig.~\ref{fig:internal}(c)), suggesting a $2{:}1$ modal relation. These observations suggest that the drive-dependent reduction in apparent $Q$ may arise from coupling-mediated energy transfer rather than purely single-mode Duffing nonlinearity. Although the drive-dependent measurements are not displacement-calibrated, the observation of such nonlinear features at the lowest applied drive levels is consistent with the low-pretension scaling.

Since internal resonance requires the participation of at least two modes, we next examine the broader mode structure of MAC nanodrums. Representative spectra of monolayer and multilayer nanodrums are shown in Fig.~\ref{fig:mode}. The spectra contain multiple vibrational modes, and the thin vertical lines mark the mode ratios expected for an ideal tension-dominated circular membrane as reference guides~\cite{leissa_vibrations_2011}. For monolayer MAC (Fig.~\ref{fig:mode}(a)), the measured mode positions deviate substantially from these ideal ratios. These deviations may arise from nonuniform residual tension, membrane morphology, local mass loading, elastic heterogeneity, residual bending rigidity, or static out-of-plane deformation~\cite{davidovikj_nonlinear_2017,kaisar_nonlinear_2022,sarafraz_quantifying_2024}. In monolayer MAC, these perturbations are expected to be amplified by the combination of low pretension and intrinsic structural heterogeneity of the amorphous network. This low-pretension regime may enhance the sensitivity of the dynamics to geometric nonlinearities, static sagging, and spatially nonuniform stress. In comparison, the 5-nm-thick nanodrums show more regular membrane-like spectra (Fig.~\ref{fig:mode}(b)), with fundamental resonance frequencies in the range 19--26~MHz and an estimated pretension of approximately $0.1~\mathrm{N\,m^{-1}}$, comparable to values reported for graphene membranes~\cite{davidovikj_nonlinear_2017,davidovikj_static_2017}. This suggests that increasing the membrane thickness can average out local structural variations, thereby restoring more conventional membrane behavior. However, the present experiment alone does not allow isolating the influence of thickness from the substantially higher pretension of these samples, nor from potential differences in morphology or mass distribution.

\section*{Conclusions}\label{sec3}

In conclusion, we have demonstrated that suspended monolayer amorphous carbon can operate as an atomically thin nanomechanical resonator. Using optothermal actuation and interferometric readout, we resolved Brownian motion, driven linear response, nonlinear dynamics, and multimode spectra in MAC nanodrums. The measured resonance frequencies indicate low effective pretension in monolayer devices, placing them in a regime where geometric nonlinearities, stress heterogeneity, and mode coupling can become important at comparatively low drive powers. Consistently, the monolayer mode spectra deviate from the ratios expected for an ideal tension-dominated circular membrane. At larger drive amplitudes, the resonators enter a rich nonlinear regime in which amplitude-dependent stiffness, nonlinear damping, parametric excitation, and intermodal coupling all contribute to the observed response. These results establish MAC as a mechanically robust system for studying how amorphous structure and low effective tension influence membrane dynamics in the two-dimensional limit. Future work correlating mechanical response with local morphology, thickness, mass loading, and stress fields should clarify the role of structural disorder in shaping dissipation and nonlinear mode coupling in atomically thin amorphous membranes~\cite{davidovikj_nonlinear_2017, kaisar_nonlinear_2022,sarafraz_quantifying_2024}.

\subsection*{Author Contributions}
L.J., F.A., and M.\v{S} conceptualised the project. L.J., A.M.P., F.A., and M.\v{S} developed the methodology. L.J., A.M.P., H.Z., P.M., N.A.M., and A.K.G. performed experimental investigations. L.J. and A.M.P. performed the formal data analysis. H.Z., P.M., and N.A.M. fabricated samples. L.J., A.K.G., P.M., and H.Z. characterised the material. A.K.G., C.T.T., B.\"O., F.A., and M.\v{S} supervised the project. L.J., B.\"O., and F.A. acquired the funding. The paper was jointly written by L.J., M.\v{S}., A.M.P. and F.A., with a main contribution from L.J. All authors discussed the results and commented on the paper.

\subsection*{Acknowledgments}
L.J. acknowledges support from the ERDF Postdoctoral Research Project No.~1.1.1.9/LZP/1/24/138. F.A. acknowledges funding from the European Union through the ERC Consolidator Grant NCANTO, Grant Agreement No.~101125458. Views and opinions expressed are, however, those of the authors only and do not necessarily reflect those of the European Union or the European Research Council. Neither the European Union nor the granting authority can be held responsible for them. B.\"O. acknowledges support from the National Research Foundation, Prime Minister's Office, Singapore, under its Competitive Research Programme, CRP Award No.~NRF-CRP22-2019-008, and from the Ministry of Education, Singapore, under its Research Centre of Excellence award to the Institute for Functional Intelligent Materials (I-FIM), Project No.~EDUNC-33-18-279-V12.

\subsection*{Conflicts of Interest}

The authors declare no conflicts of interest.

\subsection*{Data Availability Statement}

The data that support the findings of this study are available from the corresponding author upon reasonable request.

\bibliographystyle{naturemag}

\begin{thebibliography}{10}
\expandafter\ifx\csname url\endcsname\relax
  \def\url#1{\texttt{#1}}\fi
\expandafter\ifx\csname urlprefix\endcsname\relax\def\urlprefix{URL }\fi
\providecommand{\bibinfo}[2]{#2}
\providecommand{\eprint}[2][]{\url{#2}}

\bibitem{lemme_nanoelectromechanical_2020_fix}
\bibinfo{author}{Lemme, M.~C.} \emph{et~al.}
\newblock \bibinfo{title}{Nanoelectromechanical sensors based on suspended {2D} materials}.
\newblock \emph{\bibinfo{journal}{Research}} \textbf{\bibinfo{volume}{\,2020}}, \bibinfo{pages}{8748602} (\bibinfo{year}{2020}).

\bibitem{steeneken_dynamics_2021}
\bibinfo{author}{Steeneken, P.~G.}, \bibinfo{author}{Dolleman, R.~J.}, \bibinfo{author}{Davidovikj, D.}, \bibinfo{author}{Alijani, F.} \& \bibinfo{author}{van~der Zant, H. S.~J.}
\newblock \bibinfo{title}{Dynamics of {2D} material membranes}.
\newblock \emph{\bibinfo{journal}{2D Mater.}} \textbf{\bibinfo{volume}{8}}, \bibinfo{pages}{042001} (\bibinfo{year}{2021}).

\bibitem{davidovikj_nonlinear_2017}
\bibinfo{author}{Davidovikj, D.} \emph{et~al.}
\newblock \bibinfo{title}{Nonlinear dynamic characterization of two-dimensional materials}.
\newblock \emph{\bibinfo{journal}{Nat. Commun.}} \textbf{\bibinfo{volume}{8}}, \bibinfo{pages}{1253} (\bibinfo{year}{2017}).

\bibitem{keskekler_tuning_2021}
\bibinfo{author}{Keşkekler, A.} \emph{et~al.}
\newblock \bibinfo{title}{Tuning nonlinear damping in graphene nanoresonators by parametric–direct internal resonance}.
\newblock \emph{\bibinfo{journal}{Nat. Commun.}} \textbf{\bibinfo{volume}{12}}, \bibinfo{pages}{1099} (\bibinfo{year}{2021}).

\bibitem{kaisar_nonlinear_2022}
\bibinfo{author}{Kaisar, T.}, \bibinfo{author}{Lee, J.}, \bibinfo{author}{Li, D.}, \bibinfo{author}{Shaw, S.~W.} \& \bibinfo{author}{Feng, P. X.-L.}
\newblock \bibinfo{title}{Nonlinear stiffness and nonlinear damping in atomically thin {MoS$_2$} nanomechanical resonators}.
\newblock \emph{\bibinfo{journal}{Nano Lett.}} \textbf{\bibinfo{volume}{22}}, \bibinfo{pages}{9831--9838} (\bibinfo{year}{2022}).

\bibitem{Wrinkles_ArjmandiTash2026}
\bibinfo{author}{Arjmandi-Tash, H.} \emph{et~al.}
\newblock \bibinfo{title}{Mechanical reinforcement of graphene via wrinkling}.
\newblock \emph{\bibinfo{journal}{npj {2D} Mater. Appl.}} \textbf{\bibinfo{volume}{10}}, \bibinfo{pages}{65} (\bibinfo{year}{2026}).

\bibitem{liu_enhanced_2024}
\bibinfo{author}{Liu, H.} \emph{et~al.}
\newblock \bibinfo{title}{Enhanced sensitivity and tunability of thermomechanical resonance near the buckling bifurcation}.
\newblock \emph{\bibinfo{journal}{2D Mater.}} \textbf{\bibinfo{volume}{11}}, \bibinfo{pages}{025028} (\bibinfo{year}{2024}).

\bibitem{kartal_frequency_2026_fix}
\bibinfo{author}{Kartal, E.} \emph{et~al.}
\newblock \bibinfo{title}{Frequency stability of graphene nonlinear parametric oscillator}.
\newblock \emph{\bibinfo{journal}{Nano Lett.}} \textbf{\bibinfo{volume}{26}}, \bibinfo{pages}{7501--7508} (\bibinfo{year}{2026}).

\bibitem{zheng_hexagonal_2017}
\bibinfo{author}{Zheng, X.-Q.}, \bibinfo{author}{Lee, J.} \& \bibinfo{author}{Feng, P. X.-L.}
\newblock \bibinfo{title}{Hexagonal boron nitride nanomechanical resonators with spatially visualized motion}.
\newblock \emph{\bibinfo{journal}{Microsyst. Nanoeng.}} \textbf{\bibinfo{volume}{3}}, \bibinfo{pages}{1--8} (\bibinfo{year}{2017}).

\bibitem{siskins_magnetic_2020}
\bibinfo{author}{Šiškins, M.} \emph{et~al.}
\newblock \bibinfo{title}{Magnetic and electronic phase transitions probed by nanomechanical resonators}.
\newblock \emph{\bibinfo{journal}{Nat. Commun.}} \textbf{\bibinfo{volume}{11}}, \bibinfo{pages}{2698} (\bibinfo{year}{2020}).

\bibitem{liu_tuning_2023_fix}
\bibinfo{author}{Liu, H.} \emph{et~al.}
\newblock \bibinfo{title}{Tuning heat transport in graphene by tension}.
\newblock \emph{\bibinfo{journal}{Phys. Rev. B}} \textbf{\bibinfo{volume}{108}}, \bibinfo{pages}{L081401} (\bibinfo{year}{2023}).

\bibitem{tsaturyan_ultracoherent_2017}
\bibinfo{author}{Tsaturyan, Y.}, \bibinfo{author}{Barg, A.}, \bibinfo{author}{Polzik, E.~S.} \& \bibinfo{author}{Schliesser, A.}
\newblock \bibinfo{title}{Ultracoherent nanomechanical resonators via soft clamping and dissipation dilution}.
\newblock \emph{\bibinfo{journal}{Nat. Nanotechnol.}} \textbf{\bibinfo{volume}{12}}, \bibinfo{pages}{776--783} (\bibinfo{year}{2017}).

\bibitem{xu_high-strength_2024}
\bibinfo{author}{Xu, M.} \emph{et~al.}
\newblock \bibinfo{title}{High-strength amorphous silicon carbide for nanomechanics}.
\newblock \emph{\bibinfo{journal}{Adv. Mater.}} \textbf{\bibinfo{volume}{36}}, \bibinfo{pages}{2306513} (\bibinfo{year}{2024}).

\bibitem{cartamil-bueno_high-quality-factor_2015}
\bibinfo{author}{Cartamil-Bueno, S.~J.} \emph{et~al.}
\newblock \bibinfo{title}{High-quality-factor tantalum oxide nanomechanical resonators by laser oxidation of {TaSe$_2$}}.
\newblock \emph{\bibinfo{journal}{Nano Res.}} \textbf{\bibinfo{volume}{8}}, \bibinfo{pages}{2842--2849} (\bibinfo{year}{2015}).

\bibitem{zhang_vibrational_2015}
\bibinfo{author}{Zhang, X.} \emph{et~al.}
\newblock \bibinfo{title}{Vibrational modes of ultrathin carbon nanomembrane mechanical resonators}.
\newblock \emph{\bibinfo{journal}{Appl. Phys. Lett.}} \textbf{\bibinfo{volume}{106}}, \bibinfo{pages}{063107} (\bibinfo{year}{2015}).

\bibitem{toh_synthesis_2020}
\bibinfo{author}{Toh, C.-T.} \emph{et~al.}
\newblock \bibinfo{title}{Synthesis and properties of free-standing monolayer amorphous carbon}.
\newblock \emph{\bibinfo{journal}{Nature}} \textbf{\bibinfo{volume}{577}}, \bibinfo{pages}{199--203} (\bibinfo{year}{2020}).

\bibitem{tian_disorder-tuned_2023}
\bibinfo{author}{Tian, H.} \emph{et~al.}
\newblock \bibinfo{title}{Disorder-tuned conductivity in amorphous monolayer carbon}.
\newblock \emph{\bibinfo{journal}{Nature}} \textbf{\bibinfo{volume}{615}}, \bibinfo{pages}{56--61} (\bibinfo{year}{2023}).

\bibitem{will_high_2017}
\bibinfo{author}{Will, M.} \emph{et~al.}
\newblock \bibinfo{title}{High quality factor graphene-based two-dimensional heterostructure mechanical resonator}.
\newblock \emph{\bibinfo{journal}{Nano Lett.}} \textbf{\bibinfo{volume}{17}}, \bibinfo{pages}{5950--5955} (\bibinfo{year}{2017}).

\bibitem{MonolayerMAC_Zhang2026}
\bibinfo{author}{Zhang, H.} \emph{et~al.}
\newblock \bibinfo{title}{Breaking the 2‐nm barrier in hard disk drives using monolayer amorphous carbon overcoats}.
\newblock \emph{\bibinfo{journal}{Adv. Mater.}} \textbf{\bibinfo{volume}{38}}, \bibinfo{pages}{e19149} (\bibinfo{year}{2026}).

\bibitem{Thickness_Shearer2016}
\bibinfo{author}{Shearer, C.~J.}, \bibinfo{author}{Slattery, A.~D.}, \bibinfo{author}{Stapleton, A.~J.}, \bibinfo{author}{Shapter, J.~G.} \& \bibinfo{author}{Gibson, C.~T.}
\newblock \bibinfo{title}{Accurate thickness measurement of graphene}.
\newblock \emph{\bibinfo{journal}{Nanotechnology}} \textbf{\bibinfo{volume}{27}}, \bibinfo{pages}{125704} (\bibinfo{year}{2016}).

\bibitem{davidovikj_visualizing_2016}
\bibinfo{author}{Davidovikj, D.} \emph{et~al.}
\newblock \bibinfo{title}{Visualizing the motion of graphene nanodrums}.
\newblock \emph{\bibinfo{journal}{Nano Lett.}} \textbf{\bibinfo{volume}{16}}, \bibinfo{pages}{2768--2773} (\bibinfo{year}{2016}).

\bibitem{blake_making_2007}
\bibinfo{author}{Blake, P.} \emph{et~al.}
\newblock \bibinfo{title}{Making graphene visible}.
\newblock \emph{\bibinfo{journal}{Appl. Phys. Lett.}} \textbf{\bibinfo{volume}{91}}, \bibinfo{pages}{063124} (\bibinfo{year}{2007}).

\bibitem{dolleman_amplitude_2017}
\bibinfo{author}{Dolleman, R.~J.}, \bibinfo{author}{Davidovikj, D.}, \bibinfo{author}{van~der Zant, H. S.~J.} \& \bibinfo{author}{Steeneken, P.~G.}
\newblock \bibinfo{title}{Amplitude calibration of {2D} mechanical resonators by nonlinear optical transduction}.
\newblock \emph{\bibinfo{journal}{Appl. Phys. Lett.}} \textbf{\bibinfo{volume}{111}}, \bibinfo{pages}{253104} (\bibinfo{year}{2017}).

\bibitem{wang_strong_2008}
\bibinfo{author}{Wang, X.}, \bibinfo{author}{Chen, Y.~P.} \& \bibinfo{author}{Nolte, D.~D.}
\newblock \bibinfo{title}{Strong anomalous optical dispersion of graphene: complex refractive index measured by picometrology}.
\newblock \emph{\bibinfo{journal}{Opt. Express}} \textbf{\bibinfo{volume}{16}}, \bibinfo{pages}{22105--22112} (\bibinfo{year}{2008}).

\bibitem{barton_photothermal_2012}
\bibinfo{author}{Barton, R.~A.} \emph{et~al.}
\newblock \bibinfo{title}{Photothermal self-oscillation and laser cooling of graphene optomechanical systems}.
\newblock \emph{\bibinfo{journal}{Nano Lett.}} \textbf{\bibinfo{volume}{12}}, \bibinfo{pages}{4681--4686} (\bibinfo{year}{2012}).

\bibitem{barton_high_2011}
\bibinfo{author}{Barton, R.~A.} \emph{et~al.}
\newblock \bibinfo{title}{High, size-dependent quality factor in an array of graphene mechanical resonators}.
\newblock \emph{\bibinfo{journal}{Nano Lett.}} \textbf{\bibinfo{volume}{11}}, \bibinfo{pages}{1232--1236} (\bibinfo{year}{2011}).

\bibitem{leissa_vibrations_2011}
\bibinfo{author}{Leissa, A.~W.} \& \bibinfo{author}{Qatu, M.~S.}
\newblock \emph{\bibinfo{title}{Vibration of Continuous Systems}} (\bibinfo{publisher}{McGraw-Hill}, \bibinfo{year}{2011}), \bibinfo{edition}{1} edn.

\bibitem{bunch_impermeable_2008}
\bibinfo{author}{Bunch, J.~S.} \emph{et~al.}
\newblock \bibinfo{title}{Impermeable atomic membranes from graphene sheets}.
\newblock \emph{\bibinfo{journal}{Nano Lett.}} \textbf{\bibinfo{volume}{8}}, \bibinfo{pages}{2458--2462} (\bibinfo{year}{2008}).

\bibitem{zhang_probing_2023}
\bibinfo{author}{Zhang, P.} \emph{et~al.}
\newblock \bibinfo{title}{Probing linear to nonlinear damping in {2D} semiconductor nanoelectromechanical resonators toward a unified quality factor model}.
\newblock \emph{\bibinfo{journal}{Nano Lett.}} \textbf{\bibinfo{volume}{23}}, \bibinfo{pages}{9375--9382} (\bibinfo{year}{2023}).

\bibitem{samanta_tuning_2018}
\bibinfo{author}{Samanta, C.}, \bibinfo{author}{Arora, N.} \& \bibinfo{author}{Naik, A.~K.}
\newblock \bibinfo{title}{Tuning of geometric nonlinearity in ultrathin nanoelectromechanical systems}.
\newblock \emph{\bibinfo{journal}{Appl. Phys. Lett.}} \textbf{\bibinfo{volume}{113}}, \bibinfo{pages}{113101} (\bibinfo{year}{2018}).

\bibitem{arora_mixed_2024}
\bibinfo{author}{Arora, N.}, \bibinfo{author}{Singh, P.}, \bibinfo{author}{Kumar, R.}, \bibinfo{author}{Pratap, R.} \& \bibinfo{author}{Naik, A.}
\newblock \bibinfo{title}{Mixed nonlinear response and transition of nonlinearity in a piezoelectric membrane}.
\newblock \emph{\bibinfo{journal}{ACS Appl. Electron. Mater.}} \textbf{\bibinfo{volume}{6}}, \bibinfo{pages}{155--162} (\bibinfo{year}{2024}).

\bibitem{li_strain_2024}
\bibinfo{author}{Li, Z.} \emph{et~al.}
\newblock \bibinfo{title}{Strain engineering of nonlinear nanoresonators from hardening to softening}.
\newblock \emph{\bibinfo{journal}{Commun. Phys.}} \textbf{\bibinfo{volume}{7}}, \bibinfo{pages}{53} (\bibinfo{year}{2024}).

\bibitem{houri_demonstration_2020}
\bibinfo{author}{Houri, S.}, \bibinfo{author}{Hatanaka, D.}, \bibinfo{author}{Asano, M.} \& \bibinfo{author}{Yamaguchi, H.}
\newblock \bibinfo{title}{Demonstration of multiple internal resonances in a microelectromechanical self-sustained oscillator}.
\newblock \emph{\bibinfo{journal}{Phys. Rev. Appl.}} \textbf{\bibinfo{volume}{13}}, \bibinfo{pages}{014049} (\bibinfo{year}{2020}).

\bibitem{prasad_tunable_2023}
\bibinfo{author}{Prasad, P.}, \bibinfo{author}{Arora, N.} \& \bibinfo{author}{Naik, A.~K.}
\newblock \bibinfo{title}{Tunable nonlinear damping in {MoS$_2$} nanoresonator}.
\newblock \emph{\bibinfo{journal}{Appl. Phys. Lett.}} \textbf{\bibinfo{volume}{123}}, \bibinfo{pages}{263509} (\bibinfo{year}{2023}).

\bibitem{sarafraz_quantifying_2024}
\bibinfo{author}{Sarafraz, A.} \emph{et~al.}
\newblock \bibinfo{title}{Quantifying stress distribution in ultra-large graphene drums through mode shape imaging}.
\newblock \emph{\bibinfo{journal}{npj {2D} Mater. Appl.}} \textbf{\bibinfo{volume}{8}}, \bibinfo{pages}{45} (\bibinfo{year}{2024}).

\bibitem{davidovikj_static_2017}
\bibinfo{author}{Davidovikj, D.}, \bibinfo{author}{Scheepers, P.~H.}, \bibinfo{author}{van~der Zant, H. S.~J.} \& \bibinfo{author}{Steeneken, P.~G.}
\newblock \bibinfo{title}{Static capacitive pressure sensing using a single graphene drum}.
\newblock \emph{\bibinfo{journal}{ACS Appl. Mater. Interfaces.}} \textbf{\bibinfo{volume}{9}}, \bibinfo{pages}{43205--43210} (\bibinfo{year}{2017}).

\end{thebibliography}

\end{document}